\documentstyle[prb,aps]{revtex} 
\begin{document}
\draft
\twocolumn[\hsize\textwidth\columnwidth\hsize\csname @twocolumnfalse\endcsname  
\title{
Simulation of thermal conductivity and heat transport in solids
}
\author{C. Oligschleger$^{\dagger,}$\cite{COaddr} and J.C. 
Sch\"on$^{\ddagger}$
} 
\address{
$^{\dagger}$Institut f\"ur Algorithmen und Wissenschaftliches Rechnen, 
GMD -- For\-schungs\-zentrum Informationstechnik, D-53754 St.Augustin \\    
$^{\ddagger}$Institut f\"ur Anorganische Chemie und SFB408, 
Universit\"at Bonn, 
Gerhard-Domagk-Str.1, D-53121 Bonn
}    
\date{\today}
\maketitle

\begin{abstract}

Using molecular dynamics (MD) with classical interaction potentials  
we present calculations of thermal conductivity and heat transport
in crystals and glasses. Inducing shock waves and heat pulses  
into the systems we study the spreading of energy and temperature over the 
configurations. Phonon decay is investigated by exciting single  
modes in the structures and monitoring the time evolution of the amplitude 
using MD in a microcanonical ensemble. As examples, crystalline and amorphous
modifications of Selenium and $\rm{SiO_2}$ are considered. 
\end{abstract}
]
\pagebreak          
\section {Introduction}

Thermal conductivity and heat transport play an important role in the 
understanding of structural and dynamical differences between amorphous and 
crystalline substances; however, the underlying mechanisms are not yet 
fully understood. 
Since the low temperature experiments by Zeller and Pohl thermal 
conductivity and specific heat are known to show a universal and anomalous
behaviour in glasses \cite{Zeller}. Phenomenological models to explain these 
outstanding effects have been proposed by P.W. Anderson et al. \cite{Anderson} 
and by Phillips et al. \cite{Phillips} (TLS-model: tunneling in two-level 
systems), and an extension to somewhat higher  
temperatures has been worked out by Karpov et al. \cite{Karpov} 
in order to take into account contributions caused by anharmonic vibrations and
thermally activated hopping 
processes or relaxations in glasses (soft potential model). However, a fully 
consistent picture requires the analysis of both equilibrium, steady-state and 
non-equilibrium aspects of the transport properties of solids. 

During the last two decades non-equilibrium molecular dynamics (NEMD) has 
been
applied to study such properties, e.g. heat transport, the
evolution of shock waves, and decay of phonons.
\cite{Ladd,Hoover,HansenKlein,TsaiMacDonald} 
Based on these pioneering investigations it becomes possible to study the
properties of solids using realistic interaction potentials, enabling us to distinguish 
between the influence of specific features, e.g. the nature of 
{\it chemical} bonding, and a more universal behaviour, e.g. a more general  
{\it physical} dynamics. 
Since more than 25 years shock waves have been simulated using NEMD 
\cite{Tsai73,Ladd83} with the focus on shock wave propagation, on plastic  
deformation of solids or shock wave-induced melting. \cite{Belonoshko}  
Very recently large-scale computer simulations have been performed to gain 
insight into plastic deformations in solids induced by shock waves, where the 
influence of dislocations and defects on the plastic deformation is 
investigated using NEMD. \cite{Holian98} In these extensive 
simulations boundary effects could be excluded. 

The estimation of phonon lifetimes and their influence on transport properties
is stressed by several authors. \cite{Ladd,Allen} Ladd et al. 
have calculated the thermal conductivity for an FCC lattice using both density-
and heat-flux correlation functions. \cite{Ladd}
Allen and Feldman have investigated the thermal conductivity in silicon by 
deriving and evaluating a formula based on the Kubo and Greenwood-Kubo 
formalism \cite{Allen} representing 
the heat current operator in terms of eigenmodes determined from the 
dynamic matrix of the disordered Si-structure. 

Using MD-simulations, Michalski has studied the influence of harmonic 
and anharmonic contributions to the atomic interaction potential on thermal
conductivity and heat transport in two-dimensional (quasi-crystalline) 
structures. Furthermore, he has investigated the influence of delocalized and 
localized
modes, the latter being responsible for strong anharmonic effects in glasses.  
\cite{Michalski}

Our work is concerned with molecular dynamics simulations of thermodynamical 
steady-state and non-equilibrium transport properties in realistic covalent 
structures, where the main contributions to
the thermal conductivity $\kappa$ and the propagation of heat pulses and shock 
waves originate from the vibrational degrees of freedom. 
Our simulations for $\kappa$ are similar to those of Michalski 
\cite{Michalski}, but differ from this work by our use of three-dimensional 
structures with periodic boundary conditions in all lateral dimensions. 

One aim of our simulations is to check whether one can simulate and 
realistically 
describe thermodynamical properties with ``usual" interatomic potentials and 
structures in a realistic way, and how reliable such potentials and 
configurations are in modelling heat transport in solids. Another issue 
we want to address is the importance and influence of structural 
differences in glasses and crystals regarding heat transport. Here, the computer 
experiment mimicking the macroscopic set-up used to calculate $\kappa$, is 
complemented by the study of phonon decay and wave propagation in such solids.

In the next section, we briefly describe the systems and interaction potentials 
which we have used. 
In section \ref{methods}, we explain the methods in detail. The results and 
comparison with
experiments are presented in section \ref{results}, followed by a discussion of the 
results. 
 
\section{Example Systems}\label{Example} 

As representative covalent materials we have used Se and $\rm{SiO_2}$. 

Selenium readily forms glasses and amorphous structures. \cite{Do74}  
Several crystalline structures exist, including two ($\alpha-$ and $\beta-$)
monoclinic forms with four eight-membered rings packed differently in the unit
cell. Under standard conditions the most stable crystalline phase consists of
infinite parallel helical chains with trigonal symmetry \cite{Do74,Wy82}. We used 
this so-called trigonal (t-)Se for our investigations of crystalline 
selenium.  

The potential used to simulate Se has been described earlier. \cite{OJRS}  
The parametrization of the potential was chosen to mimic certain structural
properties
of selenium: The potential was fitted to reproduce bond lengths, angles and
bonding energies of small Se-molecules \cite{HJ91,BE80}, and to give a
reasonable description of the trigonal crystal. \cite{Wy82,HDH73,Mills} 

$\rm{SiO_2}$ exists in many different crystalline allotropes (e.g. $\alpha$-  
and $\beta$-quartz, 
high- and low-cristobalite, tridymite, keatite, coesite and stishovite) 
\cite{Wykoff} and is known to be a strong glass former. 
\cite{Angell,Wright85,Wright88} 
The atomic interaction potential was fitted 
by Vashishta et al. \cite{VKRE} in order to reproduce structural and dynamical 
properties of both crystalline and glassy phases. 

From experiment it is known that a-$\rm{SiO_2}$ has a quite high thermal
conductivity compared to others glasses \cite{Gmel81} which have typically 
thermal conductivities of the order of several $0.1 \rm{W/m K}$. 

Using these empirical potentials for Se and $\rm{SiO_2}$ glasses are generated 
by rapid MD-quenches
from well equilibrated liquids \cite{olig93,olig97} and a final quench to 0~K.

\section {Methods and Calculations} \label{methods}

With molecular dynamics (MD) one can simulate and investigate properties of  
complex systems. \cite{Cicotti,Tildesley,Hansen,Heermann,NoseKlein} 
In order to determine 
structural, dynamical and thermodynamical quantities one typically explores
correlation functions (e.g. Van-Hove-correlation to get an insight into the
radial distribution of atomic distances, and velocity-auto-correlation 
\cite{Dickey} or 
displacement-auto-correlation \cite{BA} to calculate the vibrational 
spectrum of configurations). 
Here we want to describe a more direct way to determine thermal 
properties of solids. 

\subsection{Thermal conductivity} \label{thermcond}

Experimentally the thermal conductivity is determined by measuring the
{\it stationary} heat flux necessary to maintain a temperature gradient 
generated by two heaters in the sample. \cite{Ibach} One fundamental problem in 
experiment is the thermal 
insulation of the sample against heat loss to the surroundings. Note that in 
computer experiments this problem can be avoided by imposing periodic boundary 
conditions. 

In this paper we describe a simulation which is designed to mimic the 
experimental setup. 
First we equilibrate configurations (with $N=1470$ to 23520 particles) at 
temperature $T$ for several thousand MD time steps (typically 4000 to 40000 
MDS).
In order to determine the thermal conductivity $\kappa$ of a given structure,
we select two "`contact"' layers of atoms (typically about 10$\%$ of all atoms comprising 
the structure) and fix the average temperatures in these layers at $T_L = T \pm \Delta 
T$ by scaling the atomic velocities according to the formula 
$\frac{3}{2}N k_B T = E_{kin}$. For symmetry reasons these layers are 
separated by half the box-length. In order to reduce surface effects  
we apply periodic boundary conditions in all spatial directions.  
The setup of our computer experiment is sketched in Fig. \ref{setup}.  
The atoms outside these layers are allowed to move according to Newton's 
laws. The velocities of these particles are not rescaled. 
After some time (typically several thousand MDS) a 
stationary temperature 
gradient ${2\Delta T}\over{L/2}$ has developed. To determine the temperature 
gradient, we calculate the ``local" temperature (i.e. the kinetic energy) of 
sub-layers of the structure by 
dividing the system into parallel layers of equal thickness in which we measure 
the ``local" (kinetic) temperature as described in Refs. 
\onlinecite{TsaiMacDonald,MacDonaldTsai}. 
In Fig. \ref{T_grad} we plot the temperature of the  
sub-layers versus the z-coordinate of the layers ($z\in [-L/2$, $L/2]$ 
corresponds to 
the original box, the additional points are plotted to show the 
periodic boundary conditions). The fluctuations of the mean temperatures  
in those sub-ensembles for which we do not regulate the temperature   
are about $\pm4\%$, whereas in the two layers, where the temperature is
fixed by scaling the velocity, 
the local temperatures change about $\pm2\%$ (this is the change of the local
temperature in one time 
step {\it before} scaling).
Note the clear development of a {\it linear}
temperature gradient during the simulation. 
This shows the development of a steady non-equilibrium state, which enables 
us to use Fourier's law  of heat flow 
\cite{Michalski,MacDonaldTsai,Mountain83},
where the 
heat flux $j$ necessary to maintain the temperature difference $2\Delta T$
is given by: 

\begin{equation}
j = \frac{\langle\Delta E\rangle}{A \Delta t} = -\kappa 
\frac{2\Delta T}{(L/2)}. 
\end{equation}
Here, $\Delta t$ is the time step used in the MD simulation, $A$ is the 
interface area of the sample perpendicular to the heat flux, and 
$\langle\Delta E\rangle$ is the 
average energy per time step $\Delta t$ which is added and subtracted,  
respectively, in the layers representing the heat contacts. 
The changes of the energy $\langle\Delta E\rangle$ are determined by the 
changes of the temperatures of the layers: We heat and cool the layers by 
rescaling the velocities of the particles comprising the layers, i.e. changing
the atomic velocities from ${\bf v}_i$ (before scaling) to ${\bf v}_i^{\prime}$
(after scaling). Therefore, the energy
difference is given by 
$\langle\Delta E\rangle = \langle\Delta E_{\rm kin}\rangle= 
\langle\frac{1}{2}\sum_{i=1}^{N_L}m_i({\bf v}_i^{\prime 2}-{\bf v}_i^2)
\rangle =
\frac{3}{2} N_L k_B 
\langle\delta T_L\rangle$, with 
$N_L$ the numbers of atoms in the layers (large enough to define a 
sensible local temperature) and 
$\langle\delta T_L\rangle$ the 
temperature change necessary to maintain the desired temperature 
$T_L=T \pm \Delta T$ of the layers. The amount $\langle\delta T_L\rangle$ 
is averaged over the last 20000 MDS of our simulation. To avoid quantum   
effects, which we cannot account for (e.g. tunneling in two-level-systems), we 
have chosen ``intermediate" temperatures to simulate and to measure thermal  
conductivity.

In order to test our ``computer experiment", we vary 
(a) the temperature gradients, 
(b) the layer thicknesses,
(c) the temperatures, and
(d) the system size. 

(a) 
We find that temperature differences between the contact layers ranging from 20 to 
40$\%$ of the average temperature are sufficient to reach convergence in the 
``measurement" of $\langle\Delta E\rangle$ in reasonable computer time. 
However, in typical computer 
simulations the temperature gradients are of the order $10^{10}$ K/m, 
which is orders of magnitudes higher than in experiments. Such large
gradients might have a strong 
influence on the decay of phonons. 

(b) 
We have varied $N_L$ 
from 1$\%$ to 20$\%$ of the total number of atoms $N$ in the simulation cell. 
In our experience we get fast convergence of the results by 
using about $N_L = 0.1~N$ atoms in the layers. 

(c) 
The temperatures used in our simulations range from several K upwards to 
about 400 K. At temperatures below 1 K, the physical effects in solids are 
dominated by quantum effects which our classical simulation method cannot  
describe. On the other hand, very high temperatures lead to large displacements
of the atoms in the layers, and the interfaces of the layers and the rest of 
the system roughen
considerably. Simulating Se-glass at temperatures ranging from 3.5 K to
170 K, we observe mean displacements $\langle u\rangle  = 0.6$ to 
0.8~\AA~per atom. 
In trigonal Se the mean displacement of the particles is about 0.2~\AA~per 
atom at $T = 370$ K. 

The drift in the total energy of the configurations is less than 4 parts in 
$10^4$ for the highest temperature employed and 
less than 3 parts in $10^6$ for the lowest temperature simulated.
The reason for this energy drift is the use of an 
isothermal
``sub-ensemble" in the part of the system where we fix the temperature, while 
treating the rest of the system as a microcanonical ensemble where the total 
energy is conserved. 

(d) 
From experiment one knows that in ``perfectly" crystalline but finite 
structures the thermal conductivity 
depends on the lateral dimensions of the samples. The smaller the 
crystalline sample, the smaller is the measured thermal conductivity. 
In such confined structures the mean 
free path of the
phonons is limited by the surfaces of the experimental probes which act as
scatterers for the phonons. This phenomenon is known as Casimir limit.
\cite{Casimir}
In computer simulations where one deals with small (typically nano-scale) 
structures, the influence of the system size will be even stronger than 
in experiments with sample diameter of the order of mm. 
In order to deal with this phenomenon, we have performed simulations for 
several system sizes $L$ and extrapolated the thermal conductivity for 
$L \rightarrow \infty$. 
In Fig. \ref{kappa_vs_L} we show the inverse of the thermal
conductivity versus the inverse of the layer distance for t-Se. 

\subsection{Heat pulse}

To study the dissipation and spread of kinetic energy in the system we  
induce a heat pulse into the system after it has equilibrated.
We scale the atomic velocities in a layer with $N_L$ particles
(typically 15$\%$ of the sample) 
to be much higher than the averaged velocities of the atoms outside this layer.
After this initial heat pulse the sample  is allowed to evolve freely. 
Again we calculate the
``local" temperatures for a sequence of sub-layers. \cite{MacDonaldTsai}  

The time evolution of the local temperature $T(t)$ in the different layers 
provides insights into the mechanisms underlying the energy transfer in 
the system. \cite{TsaiMacDonald} 
In particular comparing the evolution of crystalline and glassy systems  
elucidates differences in heat transport caused by the structure of the 
samples.                   
Especially the Fourier transform of $T(t)$
can be used to identify modes and vibrations responsible for heat transport.

We have applied heat pulses to amorphous Se and trigonal Se,  
both parallel and perpendicular to the helical chains.   

\subsection{Shock waves}\label{Shock waves}  

The sound velocities of the structures can be determined from the effects of 
moderate shock waves in the system. \cite{Michalski} We generate 
such a wave by shifting the atoms of one layer ($N_L = 0.02 -0.05~N$) 
from their equilibrium positions by about $2.5\%$ of the nearest neighbour 
distance in one direction and follow the development of the perturbation. 
As initial condition the atomic velocities are set to zero.  
Measuring the ``local" (kinetic) temperatures [as in 
Ref.\ \onlinecite{MacDonaldTsai}] in the sub-layers at every time step we 
derive the sound velocity from the space-time evolution of the temperature.  
We have applied this procedure to trigonal Se both parallel and perpendicular
to the chains, to $\alpha-$quartz, and to amorphous Selenium and 
$\rm{SiO_2}$. 

\subsection{Decay of modes \label{DECAY}}   

One of the central questions in the field of heat transport 
concerns the lifetime of phonons. In perfectly harmonic crystals, the
lifetime of phonons would be infinite. However, due to anharmonic contributions 
to the potential, which gain importance with increasing 
temperature,  
the mean free path of the phonons is limited by umklapp processes. 
On the other hand, at very low temperatures, when the umklapp processes are 
effectively frozen
out, the only limiting factors of thermal conductivity and of 
lifetimes/mean free 
paths of phonons are the scattering of phonons due to 
surface effects (Casimir limit) and/or 
impurities or defects of the 
structures.  
In real materials the thermal conductivity is also limited by 
electron-phonon interactions and isotope effects. \cite{Ashcroft,Schober_c5}

Clearly, the description of processes in terms of phonons depends strongly on 
whether
the exact states in anharmonic systems can be approximated by phonons,
i.e.,    
whether the phonons decay slowly enough to allow a meaningful description of the
instantaneous state of the system in terms of vibrational modes. 

The phonons, which correspond to the eigenvectors (EV) of the dynamical matrix
of the system, are determined in harmonic 
approximation by diagonalization of this matrix. \cite{Eispack} Since 
the number of 
eigenvectors ($3N$, $N$ being the number of particles in the simulation cell)
is very large, we can only 
estimate the lifetimes for a subset of the EVs. 
In order to determine the lifetime of a phonon, we excite an eigenmode 
$\vec e_j$ 
(normalized to unity) in the system
by displacing the atoms according to their contributions to the corresponding 
eigenvector, and follow the time evolution of the atomic motion using NVE-MD
where the number of atoms, the volume and the total energy are kept constant. 
The original displacement is given by: 

\begin{equation}
      \Delta \vec r(0) =   \alpha \vec e_j ,
\end{equation}

where $\alpha$ is the amplitude of the displacement vector. 

During the MD-simulation we calculate the projection of the actual 
atomic displacement onto a subspace of eigenmodes $\left\{ \vec e_m\right\}$: 

\begin{equation}
      c_{{\rm r e_m}}= \frac{\Delta \vec r(t) \cdot \vec e_m}
      {|\Delta \vec r(t)|}.
\label{Projection} 
\end{equation}

The projection on the eigenmode $\vec e_j$ will have the period 
$\tau_j=1/\nu_j$ of 
the excited eigenmode, and the amplitude will be constant as long as no 
interactions with other modes occur, i.e. as long as the mode does not decay. 
Due to the anharmonic interactions between the eigenvectors other modes will
become excited. The calculation of the expansion coefficients 
$c_{{\rm r e_m}}^2$ 
(with $m\ne j$) allows to monitor the excitation of other modes $\vec e_m$. 
With this procedure one can follow the time evolution of {\it single} phonons or
modes of the solids.
The usual dynamic structure factor $S(\vec Q,\nu)$ 
can be used to obtain information on the broadening of peaks and 
frequency shifts of typically {\it groups of modes}, but only little knowledge about 
the types of modes involved and their detailed interactions. The development 
of single modes can be extracted in the one-phonon approximation 
\cite{HansenKlein,GlydeHK}. 
Using the projection procedure, we have investigated the t-Se crystal and 
Se-glasses. 

\section {Results} \label{results}

\subsection{Thermal conductivity}

Using the algorithms described in Sec. \ref{thermcond} we have 
simulated the thermal heat conduction $\kappa_{||}$ (parallel to the chains) 
of trigonal Se. 
The limitations of the very short distance $\rm{L}$ between the layers   
kept at fixed temperatures $T_L = T \pm \Delta T$, which is computationally 
accessible, suggest an extrapolation of the distance to infinity. 
The simulation at $T = 25$ K
yields $\kappa = 0.033~\rm{W/(cm K)}$ for $N = 23520$ atoms. Extrapolating   
$\rm{L} \rightarrow \infty$ gives  $\rm{\kappa(L=\infty) = 0.072}$ W/(cm K), 
which is about a factor of 6 lower than the experimental result. \cite{HoPowell}
At $T = 90$ K even the 
largest 
simulated system with $N = 23520$ atoms deviates from the experimental 
value by more than a factor 3.8. 
In the limit of infinite system size, we find $\rm{\kappa(L=\infty)= 0.071}$ 
W/(cm K) for trigonal Se. This result is $30\%$ lower than the
experimental result. 
At $T = 370$ K we extrapolate $\rm{\kappa(L=\infty) = 0.017}$ W/(cm K) which
deviates from the experimental value ($\rm{\kappa} = 0.0538$ W/(cm K)
measured at $T = 400$ K. \cite{HoPowell} At temperatures above $T = 350$ K 
photons contribute to the thermal conductivity which lead to an increase
of $\kappa$. \cite{Gmel81}
The discrepancies may be due to some short-comings of the
potential which we use to model Se, especially the low-frequency phonons   
are described insufficiently. \cite{OJRS}  

In the case of Se-glasses, we find a much better agreement between the results
of our simulations and the experimental findings. In Fig. \ref{kappa_a-Se}
we plot the thermal conductivity versus temperature with double-logarithmic
scales. In the whole temperature range covered, the deviation between the 
theoretical values and the experimental ones
\cite{Meissner} never exceeds $20-30\%$. 
Furthermore, no significant system size effect has 
been detected in our calculations. 

Our simulations of $\rm{SiO_2}$-glasses show a different behaviour, however. 
To gain results of thermal conductivities comparable with experiment
it was necessary to construct  glasses with $\rm{L}$ up to 125~\AA.
From these we were able to extrapolate the thermal conductivity in the
limit $\rm{L \rightarrow \infty}$ as upper bound of the computed thermal 
conductivity
in $\rm{SiO_2}$-glasses. At $T = 60$ K we find $\kappa = 0.59$ W/(m K),  
a result 25$\%$ higher than the experimental value. \cite{Cahill87} 
Here one should note that the phonon spectrum calculated using the potential for
$\rm{SiO_2}$ given in Sec. \ref{Example} overestimates the low-frequency modes 
\cite{VKRE}, while
the calculated spectrum for Se underestimates the low-frequency phonons. 
\cite{OJRS}

\subsection{Heat pulse}

In a trigonal Se crystal consisting of 1470 atoms we excited in one layer 
parallel to the z-axis a heat pulse with a local temperature 
which was 5 times higher than the temperature in the rest of the system. 
The layer comprised $210$ atoms (14 chains with 15 atoms). 
We observed the spread of the energy in the system.  By symmetry 
the direction of the flow was perpendicular to the chains of the crystal.
The evolution of the ``local" temperature (i.e. the kinetic energy) of the 
layer $T_l(t)$ and of the rest of the system 
$T_r(t)$ are shown  in Fig. \ref{Toft} as dotted and solid lines, 
respectively. 
From Fig. \ref{Toft} one can deduce that the velocities of the atoms in the
layer with the induced higher temperature change with a period of about 0.15 ps
(corresponding to a frequency of 6.5 THz) and additional lower-frequency 
modulations in case of crystalline trigonal Se. 

The Fourier transforms of $T_{l,r}(t)$ are shown in Fig. \ref{spec_temp}.
Since $T \propto v^2$ the frequencies of the temperature changes are
a factor of 2 higher than the corresponding changes of the velocities.  
However, taking this factor 2 into account the power spectrum resulting from
the Fourier transform of 
$T_r(t)$ resembles in some features the usual density of states DOS.
\cite{OJRS,COdiss}
On the other hand, the power spectrum associated with the temperature
evolution of the excited layer shows strong contributions at low frequencies
(caused by the strong exponential decay of the temperature of this layer: the
``relaxation time" is about 350 MDS$\approx$0.7 ps). 
These modes reflect the ``dissipative"
character of the evolution of the (kinetic) temperature in the layer exposed 
to the heat pulse. 
Apart from this
dominating peak due to the fast decrease in temperature at the beginning 
of the MD-run,  
we find familiar contributions at the high frequency end caused by bond 
stretching modes, and a broad spectrum towards lower frequencies. 
\cite{OJRS,OS97} 

In another simulation, we excited a heat pulse in a 
layer parallel to the basal plane of t-Se. 
Again, the energy dissipation of the excited layer causes a strong peak 
to appear at the
low frequency part of the spectrum. However, we again find bond stretching 
modes clearly developed in the power spectrum. 
The high frequency part of the spectrum of excited vibrations in 
$\tilde T_r(\nu)$ 
resembles the one we calculate in the case of perpendicular heat flow, but 
some deviations in the middle of the spectrum occur. 
This part of the spectrum is assigned to librations and bond-bending
vibrations. \cite{OJRS}  

We have performed the same simulations for a Se-glass consisting of 1470
atoms. 
We observed a fast 
(exponential) ``heat relaxation" of the induced thermal energy, as in the case
of t-Se. 
Again, the power spectrum $\tilde T_{r}(\nu)$ exhibit 
some similarities with the typical vibrational DOS of a-Se. 

\subsection{Shock waves}

We have determined the sound velocities of solids from the analysis of shock 
waves both for crystals
and glasses (t-Se, $\alpha$-quartz, a-Se, and a-SiO$_2$). 
We divided the trigonal Se-crystal with $N = 2940$ atoms into 42 layers 
perpendicular to the chains and studied the time evolution of a perturbation 
induced in the system by    
displacing the atoms of one layer of the structure from their 
equilibrium positions by about 0.05~\AA~in z-direction. As described in Sec. 
\ref{Shock waves}, we monitor the development of the local temperature of the 
layers. 
The sound velocity of the system can be derived by plotting the time necessary 
for the perturbation to reach the various layers along the system.  
In Fig. \ref{time_layer} the times are plotted versus the layers along the
z-axis, yielding a sound velocity $\rm{v_z = 5690}$ \rm{m/s}. This result 
agrees well with earlier results obtained for the longitudinal
sound velocity $\rm{c_{33}=5630}$ \rm{m/s}. \cite{OJRS} 

After dividing the trigonal Se in layers oriented parallel to the chains (and
perpendicular to the x-axis) and displacing the  
atoms of one layer in x-direction, we again measured the speed of the heat 
pulse, finding an effective sound velocity in the   
x-direction $\rm{v_x = 4600}$ \rm{m/s}. Due to the 
symmetry of the trigonal
structure, this result is by a 
factor of
$\cos 30^{\circ}$ smaller than the longitudinal sound velocity $\rm{v_{11}}
= 5330$ \rm{m/s} calculated previously. \cite{OJRS} 

To calculate the sound velocity of $\alpha$-quartz, we investigated
the time evolution of a perturbation induced in the system by displacing 
the atoms of the central layer from their equilibrium 
positions by about 0.02~\AA~in z-direction. The structure comprising
$N= 2356$ atoms was divided into 24 layers perpendicular to the 
z-direction. 
Following the same procedure as before, we calculated the sound velocity 
$\rm{v_z = 8335}$ \rm{m/s}. This result agrees 
well with the longitudinal
sound velocity $\rm{c_{33}}= 8250$ \rm{m/s} obtained from the elastic constant 
$\rm{C_{33}} \approx 180~$GPa.\cite{olig}

Next we investigated both a-Se
with $N = 1470$ particles and a-$\rm{SiO_2}$ with $N=3456$ atoms. We 
divided the structures into 21 and 24 parallel layers, respectively. These 
layers were oriented perpendicular to the z-axis, and we induced a perturbation 
into the glasses by displacing all the atoms in the middle of the system  
in positive z-direction by about 0.05 and 0.02~\AA, respectively. 
As before, 
we observed the time dependence of the induced shock
wave by following the kinetic temperature evolution of the various layers. 
In both cases we did not observe any clearcut wave-like perturbation      
propagating through the system. Therefore, it was impossible to determine 
reliably, or at least to estimate the sound velocities of the glasses 
from these simulations. 
This negative outcome of the analysis of the behaviour of shock waves in glassy 
and disordered 
structures contrasts with the findings 
for the crystalline counterparts and might be traced back to structural  
differences, e.g. the lack of long-range order in amorphous structures.
Thus structural defects in glasses might cause the ``overdamping" 
of the wave, which is most likely related to the rather fast decay of phonons  
in glasses (cf. Sec. \ref{modedecay}).   

\subsection{Decay of modes} \label{modedecay}

Exciting single modes in the system allows us to study their time evolution.
We applied the procedure described in Sec. \ref{DECAY} to a subset of 
modes only,  
since the trigonal Se-crystallite chosen   
comprised $N = 1470$ atoms. In the   
low-frequency regime $(\rm{\nu \le 5 THz})$ we investigated the decay   
of n(=25) modes. To estimate the influence of temperature, we excited 
the EVs 
both for the $T = 0$ K-configuration and for a crystalline configuration 
equilibrated at $T = 60$ K. 
In Fig. \ref{Mode_exmp} we plot the projection $\rm{c_{r e_j}}$ [cf. Eq. 
\ref{Projection}] versus observation time $t$. The initially excited mode has
a frequency $\nu = 1.39$ THz and is delocalized, its participation ratio being
$p_{\nu} = 0.62$. The envelope of the projection exhibits an exponential decay
with a time-constant $\tau \approx 70$ ps. The Fourier transform 
gives insight into the frequency shift of the modes due to anharmonic effects
and into the spectral broadening of the mode. The latter effect is also
relevant for the lifetime of a mode but not so reliable due to the finite 
simulation time.  
The Fourier transform of the projection $\rm{c_{re_j}}$ shown in
Fig. \ref{Mode_exmp} leads to the spectral density plotted 
in Fig. \ref{Mode_exmp_fourier}
with a maximum at $\nu = 1.36$~THz. This shift in frequency 
means that the mode has softened with temperature.
\cite{Schober_c5}
In order to verify, whether the shift in frequency is 
due to temperature effects or is an artefact of the damping of the mode,   
we have calculated the Fourier transform of $\cos(\omega t)\exp(-t/\tau)$
with $\omega = 1.39~$THz and with $\tau \approx 7 \cdot 10^{-10},7\cdot10^{-
11},
7 \cdot 10^{-12}$s. Even for the shortest decay time $\tau$ we find a shift 
less than $10^{-4}\omega$. Thus we conclude that the observed shift in frequency
is really caused by the influence of the temperature on the modes, in 
qualitative agreement with the softening observed in experiment. 
\cite{Schober_c5} 

Due to interactions between the modes additional frequencies are 
excited in the system. In order to estimate the excitation of additional
modes, we have calculated the projections of the actual displacement on modes 
$\left\{\vec e_m\right\}$ with frequencies similar to the initially excited 
one. We sum the square of the
projections on 50 modes. After an
observation time of about 35 ps, the square of the expansion coefficient
of the initially excited mode is still larger than 0.6, and the sum for the 
energetically
similar modes is 0.3. Obviously, the atomic displacements are still 
restricted to a small subspace of EVs. Thus, only 1$\%$ of the eigenmodes of the
configuration are needed to describe the atomic motions to 
more than 90$\%$. 

\begin{figure}
\caption{ Structure factor $S(\vec Q,\nu)$ in $[0 \xi 0]$-direction for
two values of Q: solid line $\vec Q = 2\pi/a[0~0.2~0]$ and dashed line 
$\vec Q = 2\pi/a[0~0.3~0]$.} 
\label{sqw}
\end{figure}  
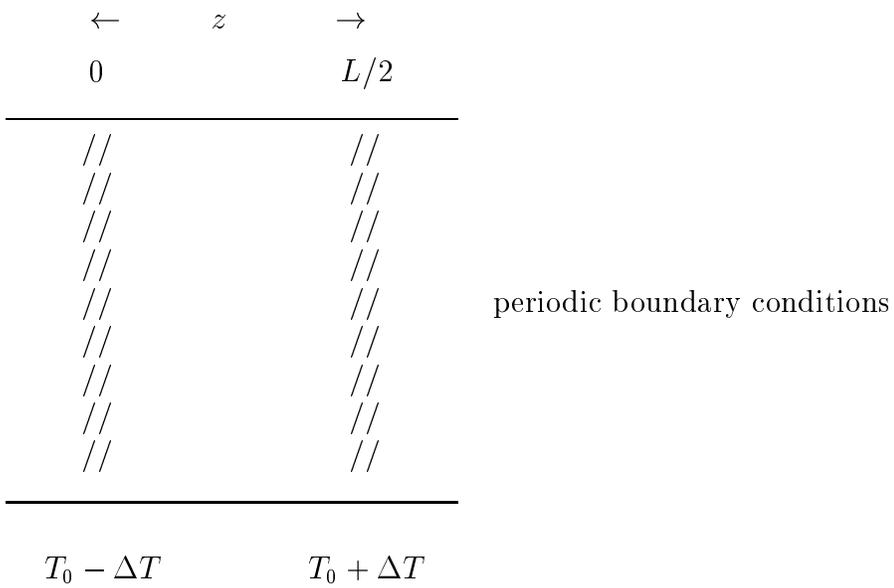
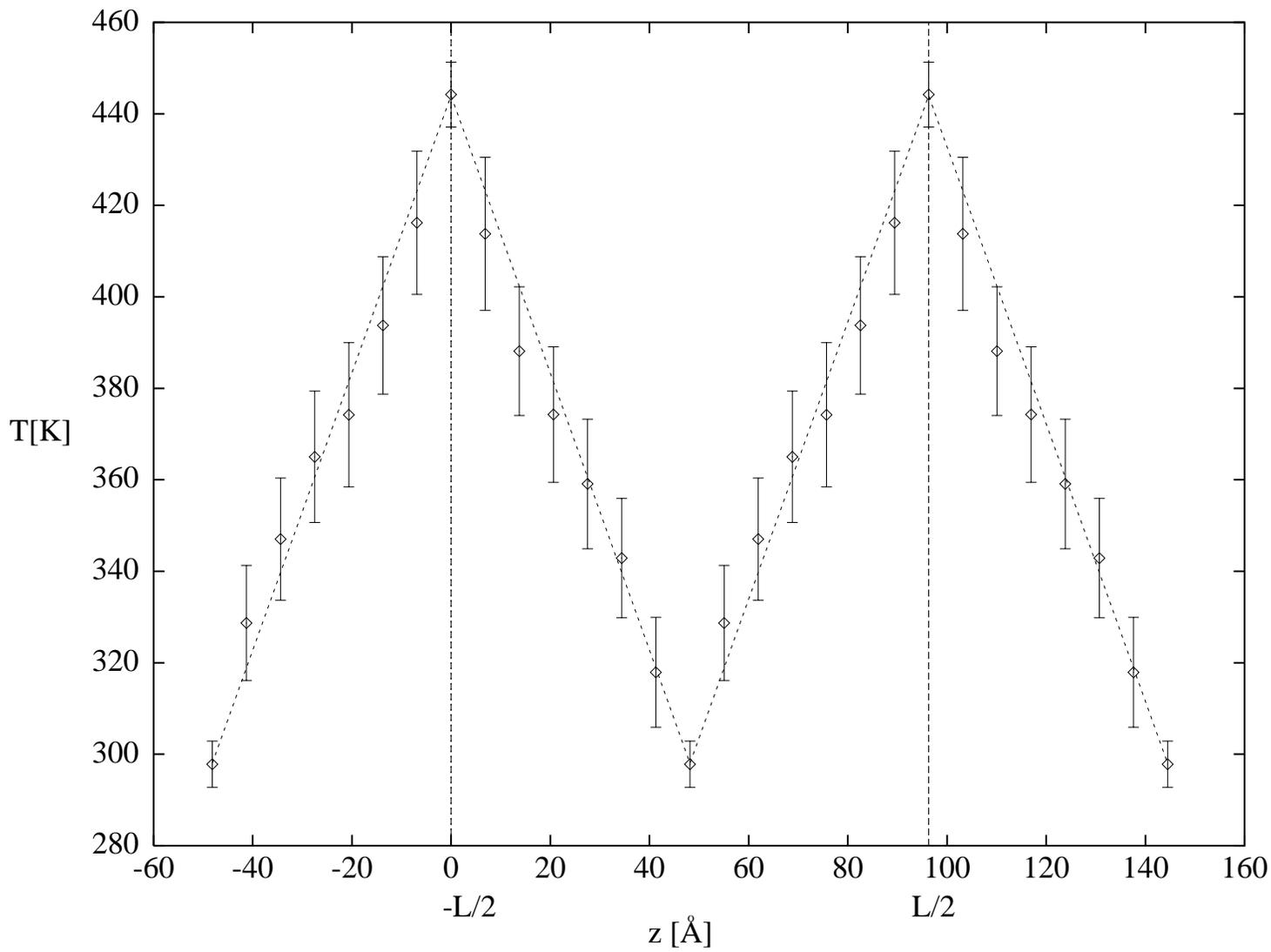
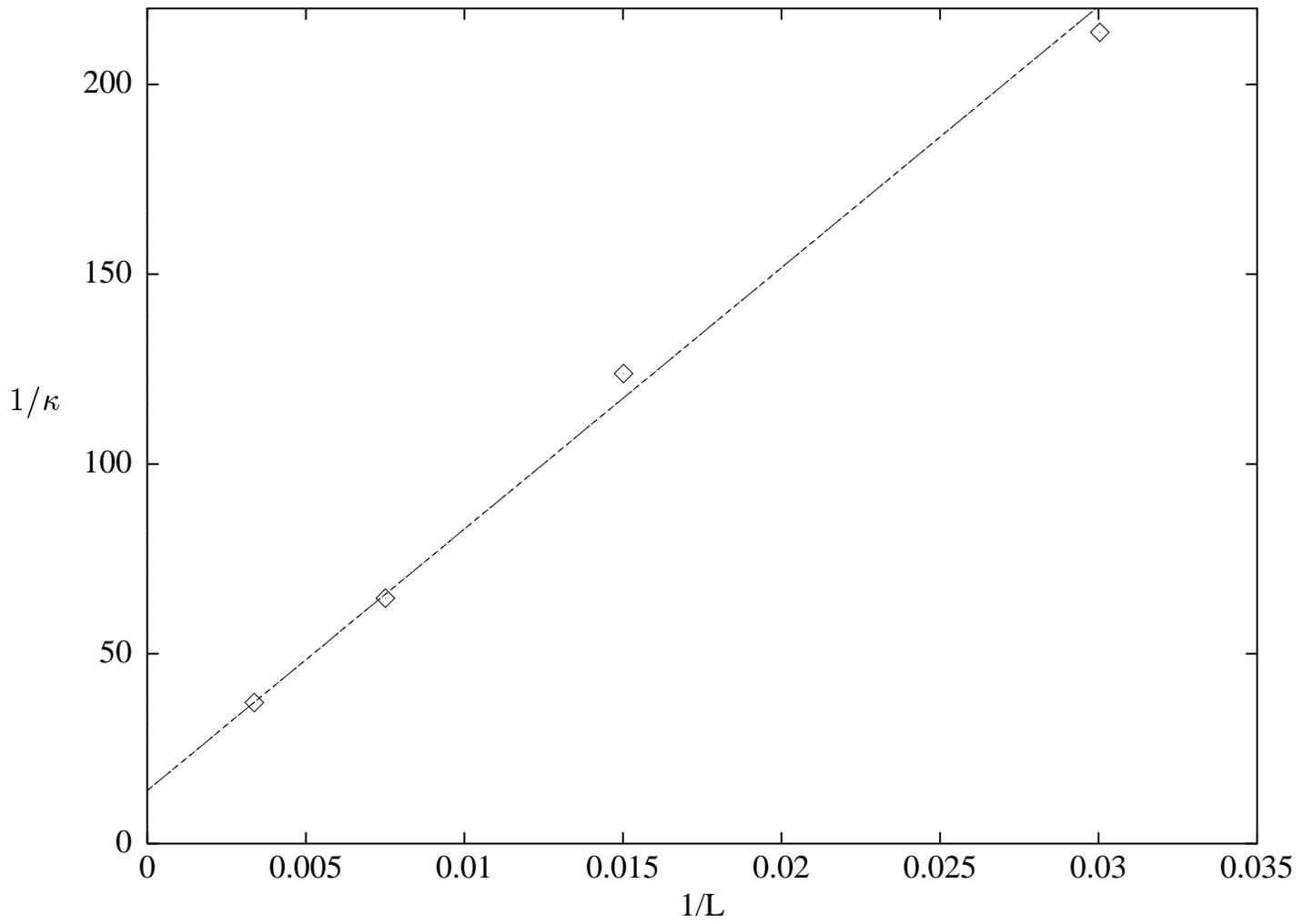
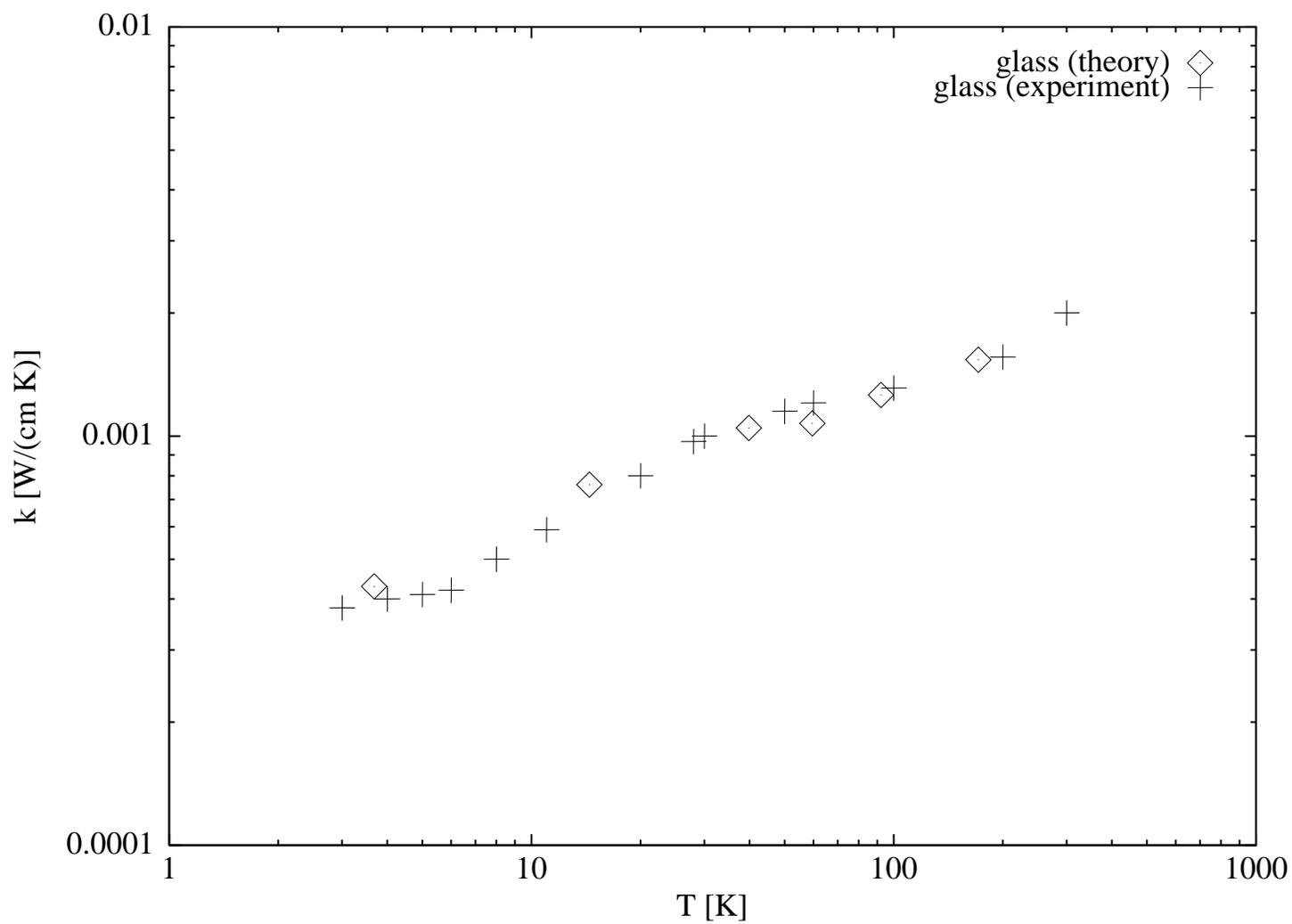
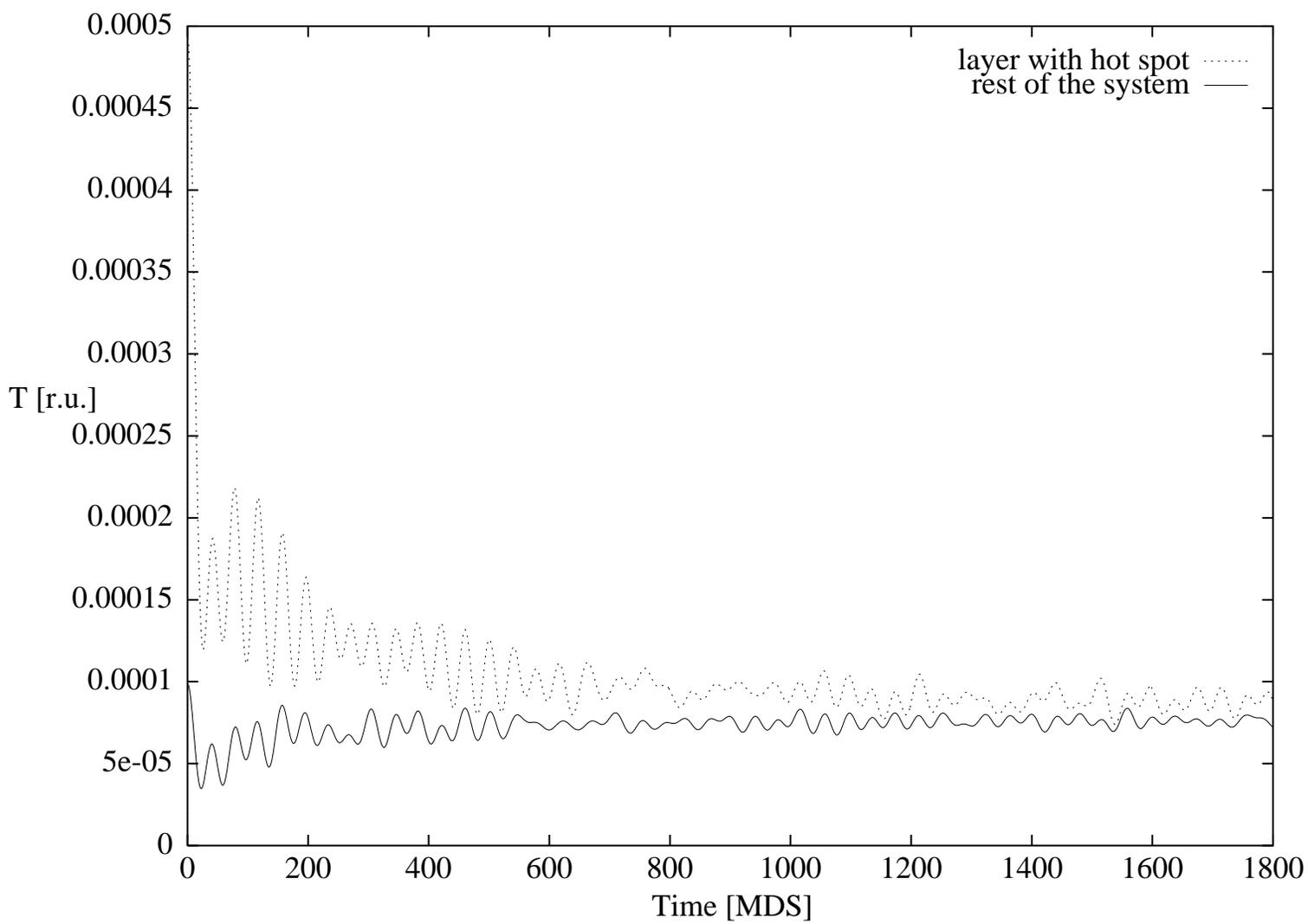
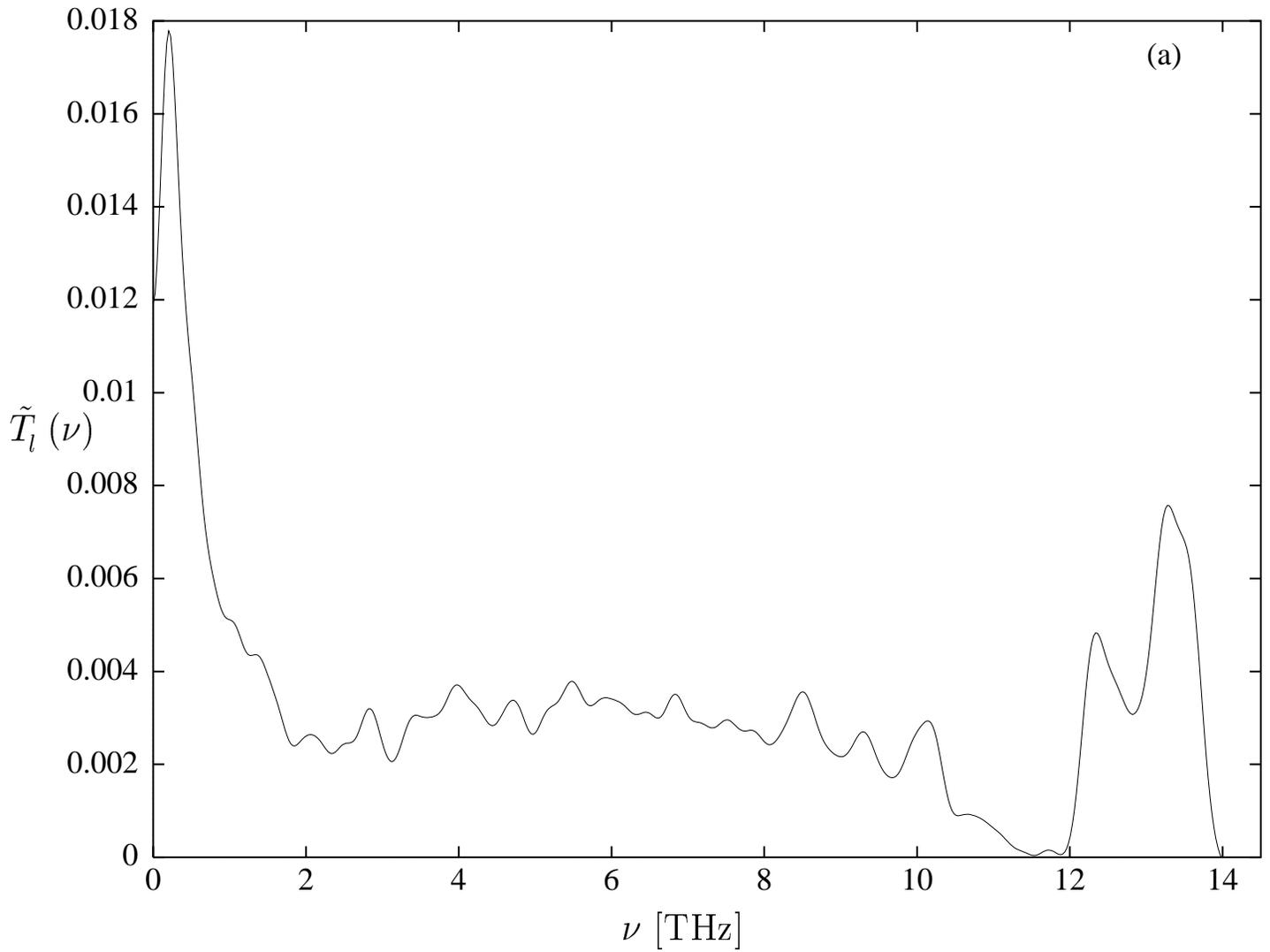
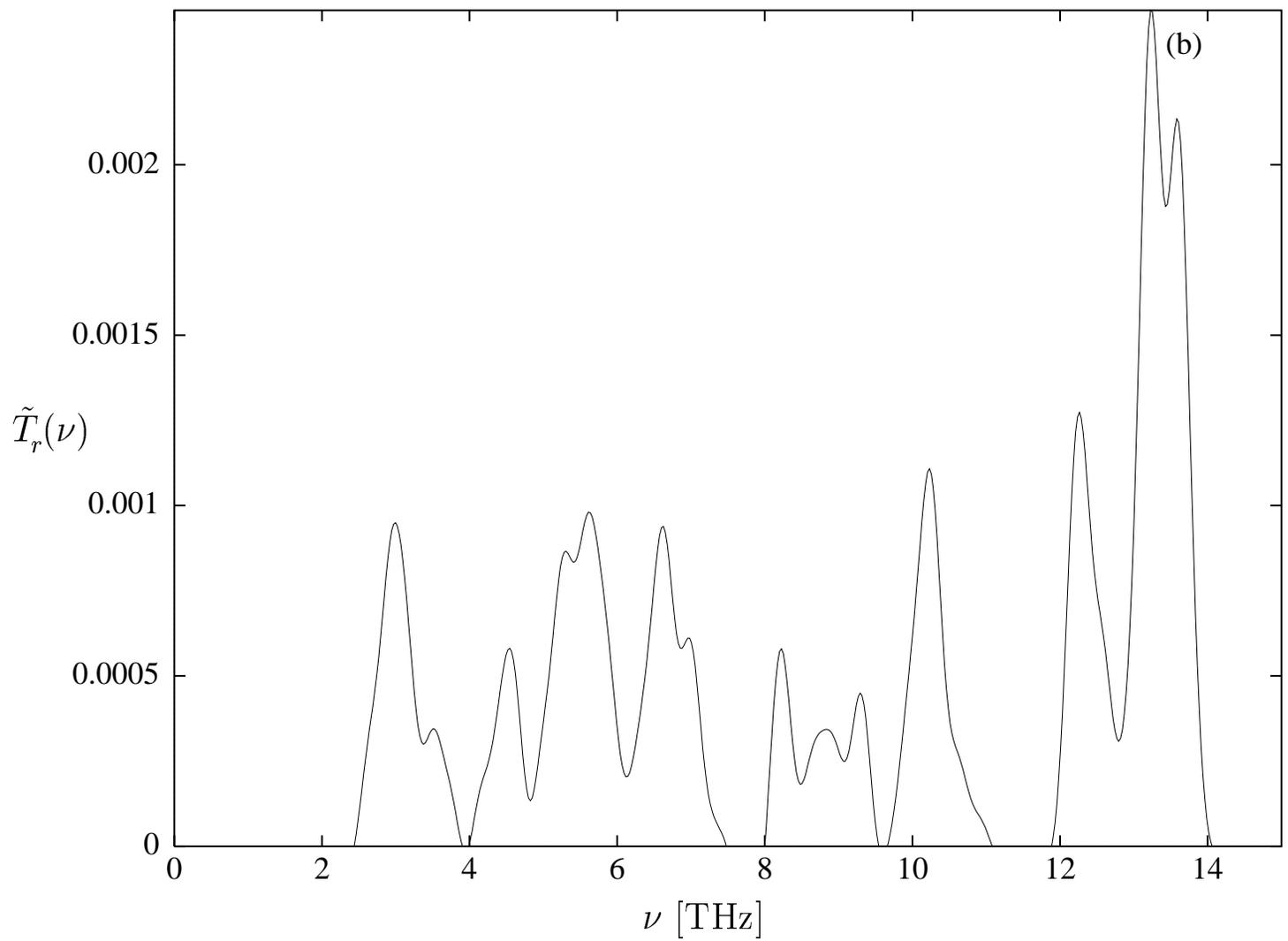
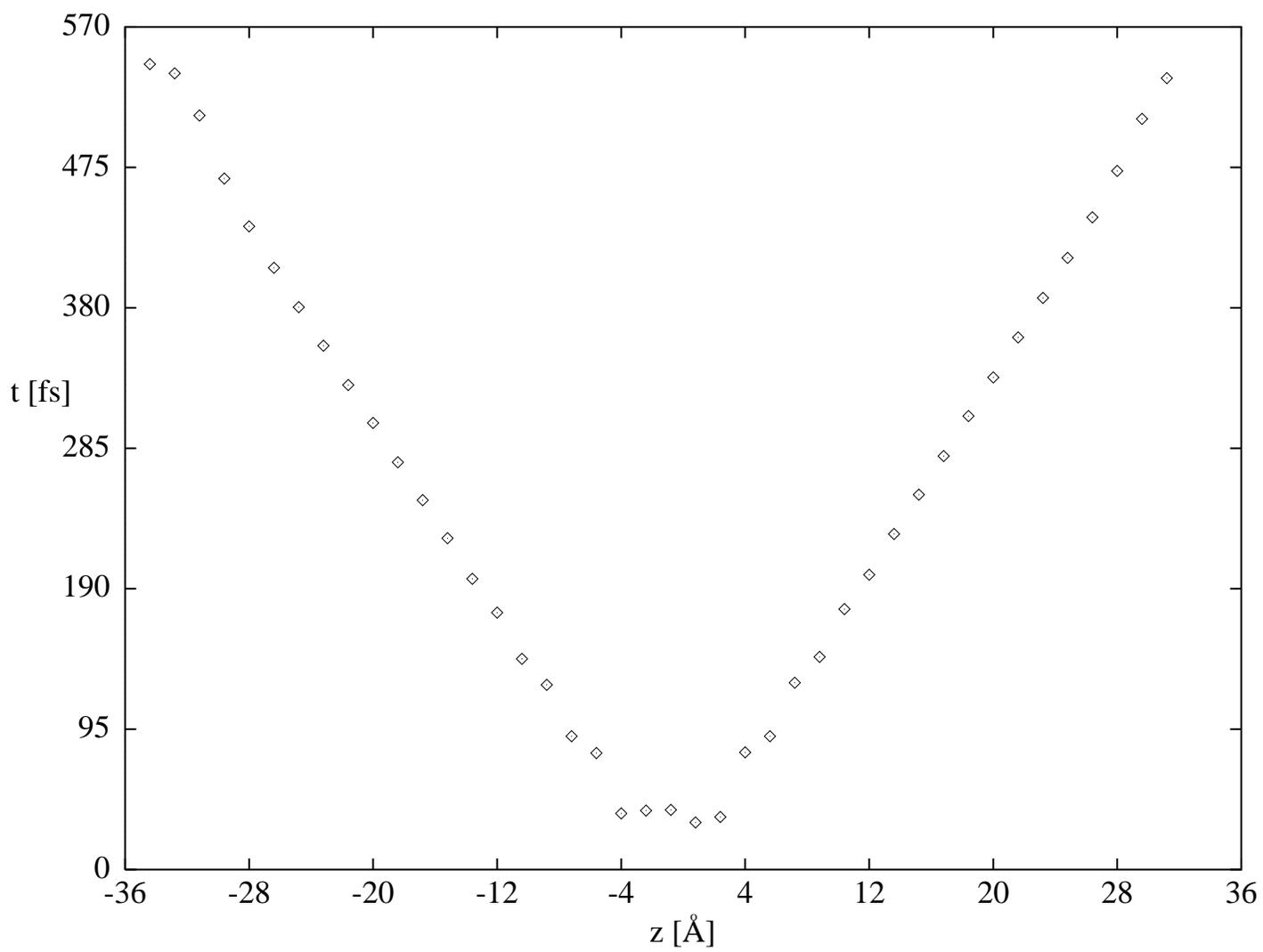
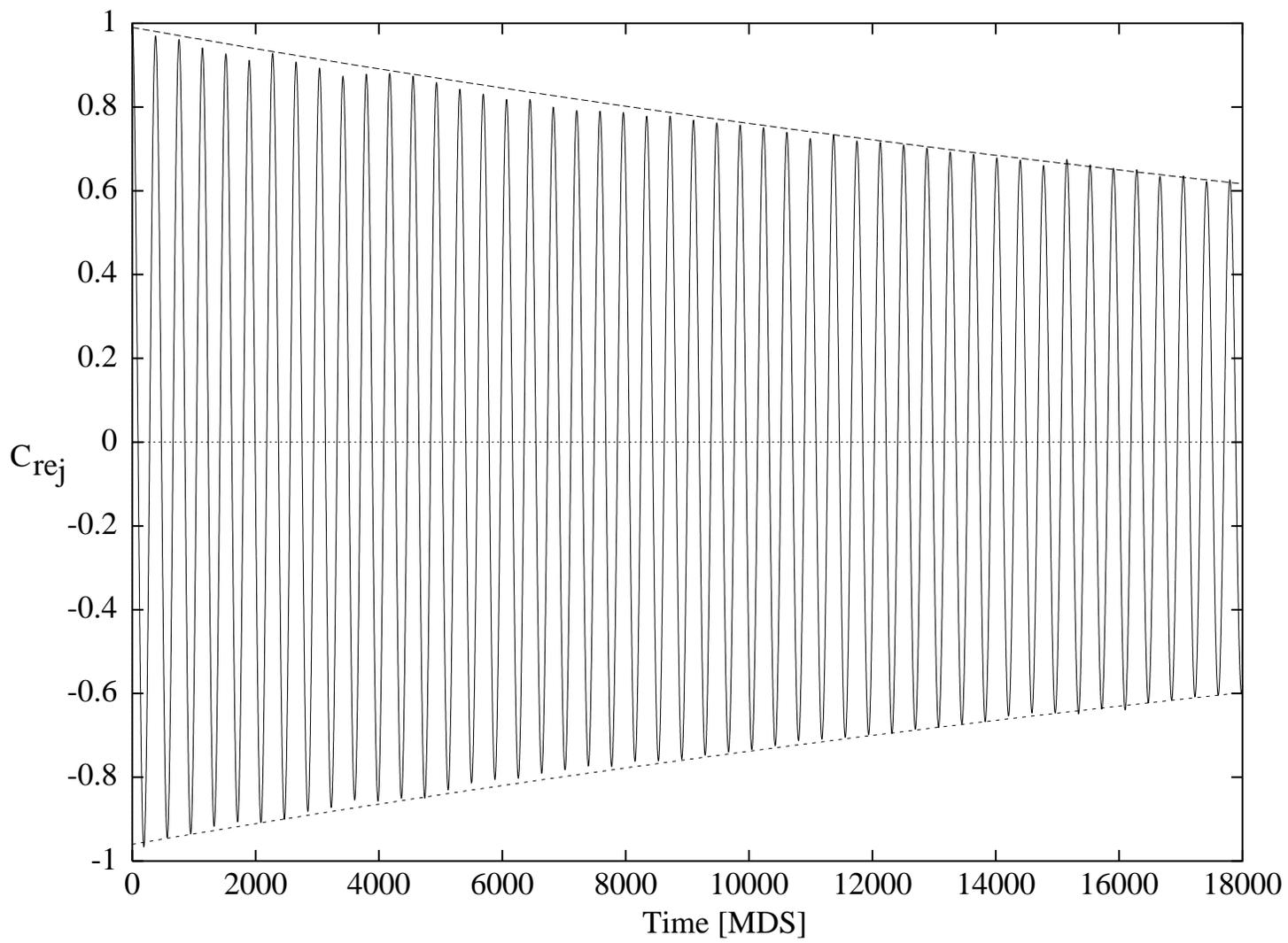
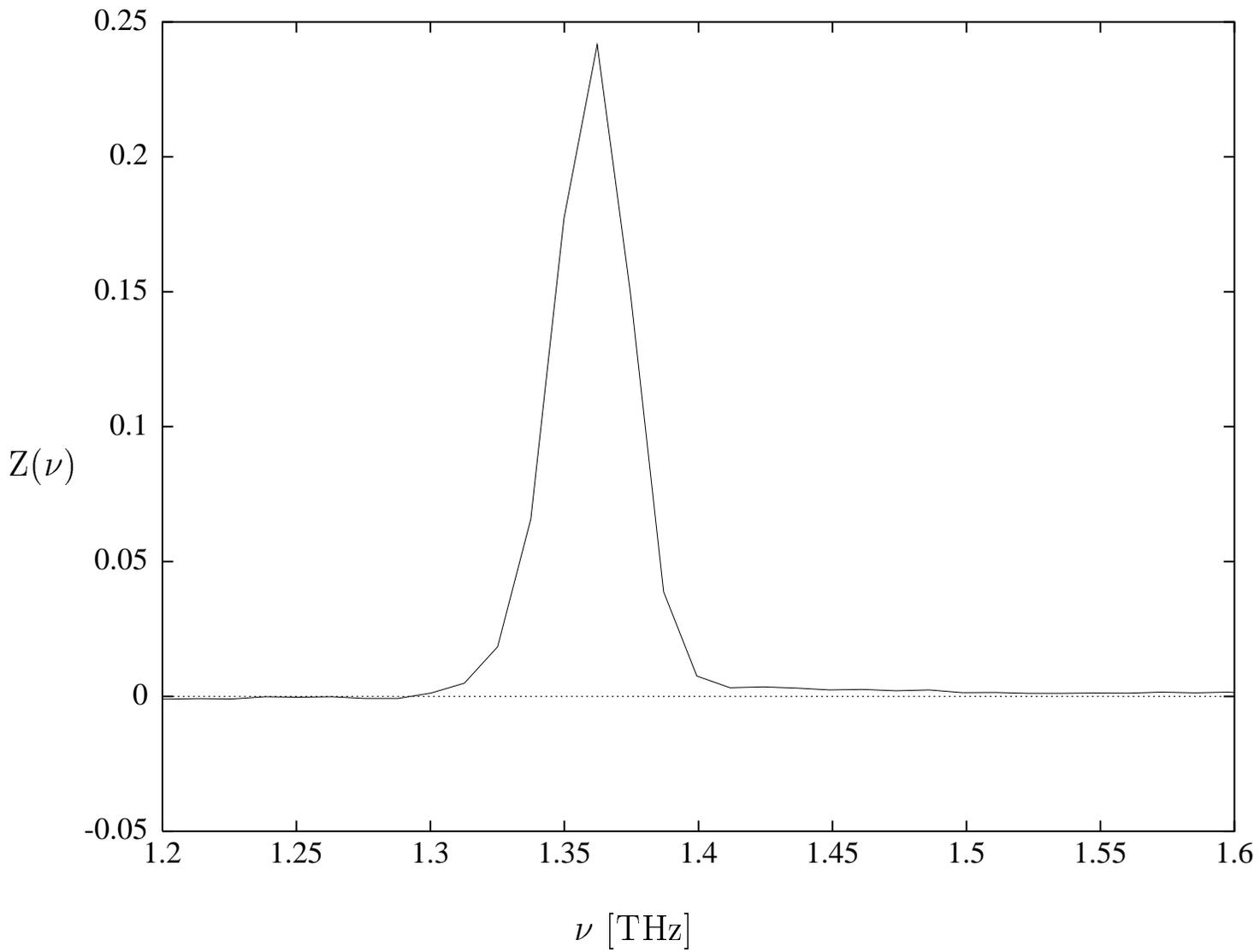
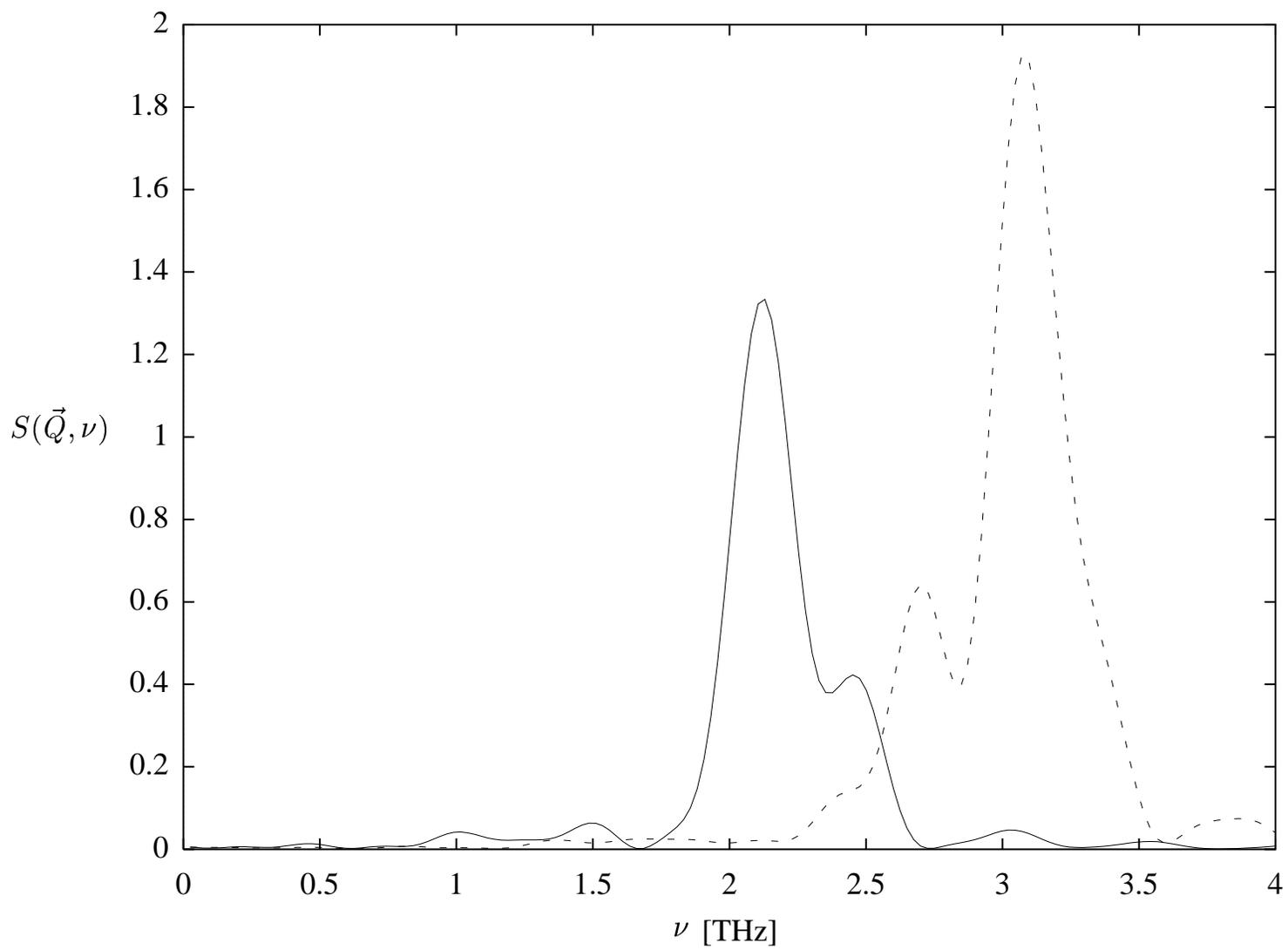

This detailed analysis can be complemented by a calculation of $S(\vec Q,\nu)$ that has the advantage to be directly comparable 
with experiment. However, in general for a given $Q$ several phonons  
contribute to the same $S(\vec Q,\nu)$. Therefore, an unambiguous assignment of 
the spectrum of $S(\vec Q,\nu)$ to {\it single} modes is not possible.
As an example we show in Fig. \ref{sqw} for two values of $Q$ in the $[0~\xi~0]$-direction 
the resulting $S(\vec Q,\nu)$. The estimate of the lifetime by FWHM (full width
half maximum) yield for the dominant phonon in  $\vec Q = 2\pi/a[0~0.2~0]$
approximately 25~ps. But due to the superposition of the peaks it is 
difficult to estimate the lifetimes of the other phonons. In particular, there are several phonons which cannot be investigated calculating 
only $S(\vec Q,\nu)$ because of this 
Q-selectivity.

In the high frequency regime we followed the time evolution of a mode just 
above 
the gap of the DOS. The eigenfrequency of this (localized) mode was $\nu = 6.2$ 
THz, and it decayed within 2 ps.
However, the simulations of extended modes in the high frequency end of the DOS 
did not show any appreciable decay of the EVs 
within the observation time. 

For glasses, the behaviour of the modes is totally different. We have 
diagonalized the
dynamical matrix of one Se-glass and calculated the EVs. In the low-frequency 
regime there exist beside extended eigenvectors a few quasi-localized 
modes, while in the high-frequency part of the spectrum nearly all modes are 
localized. 
In the regimes of low, medium and high frequencies we have performed similar 
simulations as in the case of trigonal Se. 
For a quasi-localized
mode (with a frequency $\nu = 0.31$ THz and a participation ratio 
$p_{\nu} = 0.09$) we observe a fast decay of 
the eigenmode. No clear vibration develops or can be identified in the glass. 
During the MD-run, we also checked, whether a subset of about 20 
modes with similar frequencies might be more stable. When analyzing phonons 
in the 
low-frequency regime we found that this small subset of eigenmodes (less
than 1$\%$ of the EV of the configuration) described the atomic motions to an 
extent of about 30$\%$.  
The phonons of the medium and high-frequency regime were much less stable and
decayed within several ps. 

Finally we considered  the decay of phonons in the 
non-equilibrium steady 
state, in order to estimate the influence of the strong temperature gradient
on the lifetimes of the eigenvectors. We excited some phonons in  
a t-Se crystallite with $N = 1470$ atoms, where two layers were kept at 
fixed temperatures $T_L =T \pm \Delta T$, 
and monitored the projection of the actual displacement onto
the additionally excited eigenvector. In comparison with the lifetime of the 
same mode excited in a configuration with constant temperature, the phonon 
proved to be much less stable: typically about a factor of 5. However, since we 
extrapolate to infinite system size 
keeping the temperature difference between the contacts constant, the effect of the strong temperature gradient present during the simulations should be
 largely eliminated when calculating $\rm{\kappa(L=\infty)}$. 

\section{Summary and Conclusion}

In this paper we have presented simulations of steady-state and non-equilibrium 
thermal 
heat transport in condensed
matter. As examples we used Selenium and $\rm{SiO_2}$, both in 
crystalline and amorphous modifications. 

In the crystalline structures the thermal conductivity showed a clear 
system size dependence (Casimir-limit), due to the restrictions of the mean
free path of the phonons by the lateral dimension of the simulation cell.
Extrapolation of the system size 
$\rm{L}$ to infinity yielded values of the thermal conductivity
of the same order of magnitude as the 
experimental results. Since there are no impurities in the system, the
phonons are not scattered at structural defects, vice versa this means that 
the lifetimes of the eigenmodes are limited only by anharmonic effects of the 
atomic interaction potential. 

In glasses the situation is rather different. In the case of Selenium 
glasses, 
it was sufficient to simulate structures comprising 1470 particles in the 
simulation cell to reach
convergence of the results for the thermal conductivity. This can be understood
considering that the amorphous structure, or to be more precise the state of 
disorder itself, is the reason for the anomalous and universal behaviour of
thermal heat transport in glasses. The phonons decay fast within several ps.
But nevertheless, the eigenmodes typically interact with energetically similar
modes \cite{Schober96} which means that the ``subspace" of modes is relatively
stable. Quartz-glass is known to possess extremely large mean free paths of
phonons \cite{Gmel81}; this agrees with our observation that it was necessary 
to construct glass configurations
with $L \approx 125~$\AA~in order to get results comparable  
to the experimental findings. 

It is a striking result of our work that the agreement between theory and
experiment for thermal conductivity of glasses
is better than for the corresponding crystalline phases.   
It seems that the influence of the structural disorder is so strong that 
the element-specific details of the interatomic potential are
less important than in crystalline structures. 

Complementing these steady-state computer experiments, simulations of shock 
waves and heat pulses in condensed 
matter were performed. Again one can clearly see the influence of structural order 
or disorder 
on the results and the outcome of the calculations. 

In the crystalline structures it was possible to identify waves moving through
the system. This is clearly connected to the relatively high stability of the 
modes contained in the 
induced wave packet. The behaviour of the wave packet reflects the 
elastic properties of the solid, e. g. one can calculate the sound velocity,
from which one can then  
deduce the corresponding elastic constant. 
In contrast glasses do not show  
clearly developed propagating waves after inducing a heat pulse or a shock wave
into the structure. Again, the reason for this behaviour can be found in the 
short lifetimes and mean free paths of the eigenmodes/phonons of glasses
where the disorder plays the role of scatterers for the modes. 

\section{Acknowledgments}

We are indebted to V.I. Kozub and H.R. Schober for valuable discussions and
critically reading the manuscript. 
We gratefully acknowledge funding by the Deutsche Forschungsgemeinschaft 
through the Sonderforschungsbereich 408. 
This work has been conducted within the GMD research group `Computational 
Methods in Chemistry' .
Parts of the calculations are performed on the IBM SP2 computer at the
Institute for Algorithms and Scientific Computing (SCAI), German National 
Research Center for Information Technology.   

{\it Figure captions}

\begin{figure}
\caption {Realization of the computer experiment in order to determine the    
thermal conductivity of solids. In all spatial directions we apply periodic 
boundary conditions. The shaded areas symbolize the layers where the 
temperatures are fixed.}
\label {setup}
\end{figure}
\begin{figure}
\caption {Mean temperatures in the sub-layers of the structure vs.  
z-axis.}    
\label {T_grad}
\end{figure}
\begin{figure}
\caption {The inverse thermal conductivity in $[\rm{W/cmK}]^{-1}$ for different 
system sizes plotted versus inverse system size in reduced units
at 
T = 90 K for t-Se.} 
\label {kappa_vs_L}
\end{figure} 
\begin{figure}
\caption {Thermal conductivity $\kappa [\rm{W/cmK}]$ of Se-glass plotted 
versus temperature 
T in K in a double logarithmic plot. $\diamond$: simulation, $+$: experimental 
values from Ref. \cite{Meissner}.} 
\label {kappa_a-Se} 
\end{figure} 
\begin{figure}
\caption{Evolution of the local temperature of the layer $T_l(t)$ (dotted line)
and the rest 
of the crystal $T_r(t)$ (solid line) in K vs. time in MDS after a hot spot was 
induced into a layer parallel to the chains a trigonal Selenium crystallite   
with $\rm{N = 1470}$ atoms. The typical period of temperature changes
is about 77 fs.}  
\label{Toft}
\end{figure}
\begin{figure}
\caption{ Fourier transform (power spectrum) $\tilde T_l(\nu)$ (a) and
$\tilde T_r(\nu)$ (b) of $T_{l,r}(t)$ of 
Fig. \ref{Toft}.  } 
\label{spec_temp}
\end{figure} 
\begin{figure}
\caption{ Times in fs needed for the perturbation to reach the layers are 
plotted
vs. the distance of the layers in \AA~~ along the direction of propagation. } 
\label{time_layer}
\end{figure}  
\begin{figure}
\caption{ Projection $\rm{c_{re_j}}$ of the actual displacement onto the 
initially activated eigenvector for a mode with $\nu = 1.36$ THz in the 
t-Se vs. time in MDS. The system is equilibrated at T= 60 K. 
The envelopes of the projections are two exponential functions.} 
\label{Mode_exmp}
\end{figure}  
\begin{figure}
\caption{ Fourier transform of the projection $\rm{c_{re_j}}$ shown in   
Fig. \ref{Mode_exmp}.} 
\label{Mode_exmp_fourier}
\end{figure}  
\end{document}